\documentclass[manuscript]{aastex}
\usepackage{lineno}
\usepackage{tikz}
\usepackage{amsmath}

\usepackage{graphicx}

\usepackage{pdflscape}

\slugcomment{\textbf{2025 August 20} }

\shorttitle{3I/ATLAS}
\shortauthors{Jewitt et al.}

\begin{document}

\title{Hubble Space Telescope Observations of the Interstellar Interloper 3I/ATLAS}

\author{
David Jewitt$^{1}$, Man-To Hui$^2$, Max Mutchler$^3$, Yoonyoung Kim$^1$ and Jessica Agarwal$^4$
} 
\affil{$^1$Department of Earth, Planetary and Space Sciences, UCLA, Los Angeles, CA 90095
\affil{$^2$Shanghai Astronomical Observatory,  Shanghai, China}
\affil{$^3$ Space Telescope Science Institute, Baltimore, MD 21218}
\affil{$^4$ Institute for Geophysics and Extraterrestrial Physics, TU Braunschweig, 38106 Braunschweig, Germany}
}

\email{djewitt@gmail.com}

\begin{abstract}
We present high angular resolution observations of the third known interstellar interloper, 3I/ATLAS, from the Hubble Space Telescope.   The object is clearly active at 3.8 au pre-perihelion, showing dust emitted from the hot Sun-facing side of the nucleus and a weak, radiation pressure swept tail away from the Sun.  We apply a simple model to estimate the mass loss rate in dust as   $dM/dt \sim$ 12 $a_{\mu}^{1/2}$ kg s$^{-1}$, where $a_{\mu}$ is the mean particle size in microns.  With 1 $\le a_{\mu} \le$ 100, we infer $dM/dt \sim$ 12 to 120 kg s$^{-1}$.  A fit to the surface brightness distribution of the inner coma limits the effective radius of the nucleus to be $r_n \le$ 2.8 km, assuming red geometric albedo 0.04.  Conversely, the nucleus cannot be smaller than $\sim$0.22 km in radius if its coma is supplied by sublimation of carbon monoxide, and must be larger if a less volatile molecule drives the mass loss.  
\end{abstract}


\section{INTRODUCTION}
\label{intro}

Comet C/2025 N1 (ATLAS) was discovered on UT 2025 July 1 as part of the ATLAS sky survey \citep{Den25}. Its orbit is strongly hyperbolic, with eccentricity $e$ = 6.143, inclination $i$ = 175.1\degr~and perihelion at $q$ = 1.357 au (solution reference JPL\#21, data arc from 2025 May 22 to 2025 July 28).  These parameters establish C/2025 N1 as the third known interstellar interloper, after 1I/'Oumuamua and 2I/Borisov.  Accordingly renamed as 3I/ATLAS (hereafter simply ``3I''), the orbit is the most dynamically extreme of any object yet recorded in the solar system, with a velocity at infinity of approximately 60 km s$^{-1}$.  The perihelion date is UT 2025 October 29.

Science interest in the interloper population centers on their apparent wide diversity of physical properties (1I/'Oumuamua appeared asteroid-like while 2I/Borisov was cometary), and on their flux into the planetary region of the solar system.   Both topics are relevant to understanding the origins of the interstellar interlopers, which probably lie in the forming planetary systems of other stars but which might involve tidal disruption and other processes \citep{Jew23}.  The discovery of 3I four months before perihelion when at 4.5 au presents the opportunity to study the rise of activity in response to what is almost certainly the first substantial heating event experienced by this body since its formation. The aim of the current paper is to present the first high resolution optical observations, to set a limit to the radius of the nucleus, and to obtain a preliminary characterization of the dust activity.  

\section{OBSERVATIONS}

Observations were acquired using the 2.4 m diameter Hubble Space Telescope (HST) and the WFC3 charge-coupled device camera under target-of-opportunity program GO 17830.  Images over a 6\arcmin$\times$6\arcmin~field of view were obtained at a scale 0.04\arcsec~pixel$^{-1}$.  To reach maximum sensitivity we used the F350LP filter, which has a central wavelength near 5874\AA~and a full width at half maximum (FWHM) $\sim$4800\AA.  Owing to constraints imposed by the operation of HST  in Reduced Gyro Mode, we were able to secure guided, non-sidereal observations in only a small fraction of each of two HST orbits, ultimately securing two 25 s integrations in one orbit and two 40 s integrations in the other.  A journal of observations is given in Table \ref{geometry}.

\subsection{Nucleus Size}
\begin{figure}
\epsscale{0.65}

\plotone{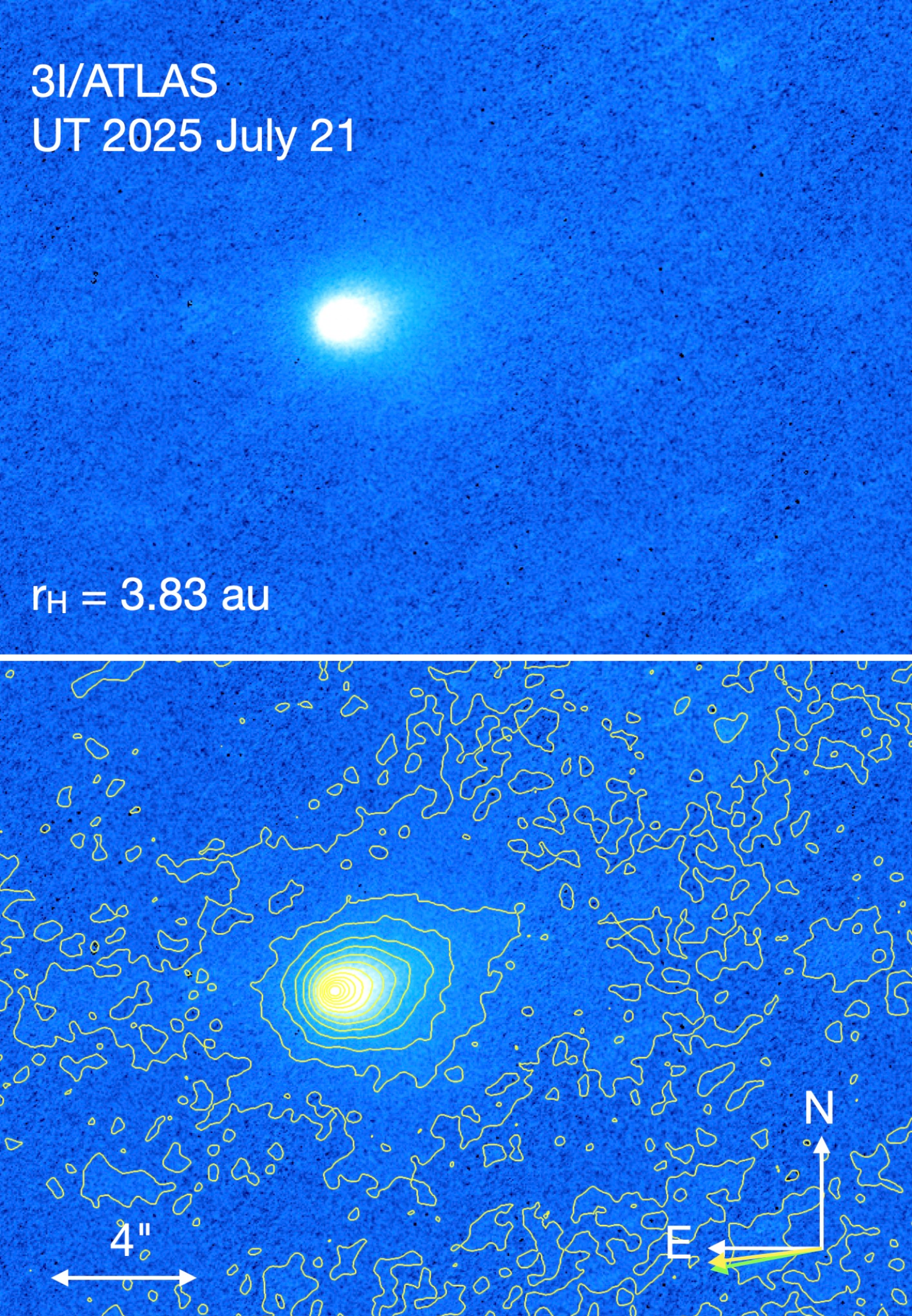}

\caption{(upper) Combined 130 s F350LP image of 3I/ATLAS showing diffuse asymmetric emission to the north west.  (lower) Same image contoured, with scale bar and direction arrows shown.  The yellow and green arrows mark, respectively, the projected negative heliocentric velocity vector and the projected anti-solar direction. Note that the bulk of the dust is sunward of the nucleus.   \label{image}}
\end{figure}

The images (Figure \ref{image}) and surface brightness profile (Figure \ref{SBplot}) show that 3I is active with no clear evidence for a central condensation that might be attributed to scattering from a nucleus.  To set a crude limit on the nucleus we first obtained photometry within a circular aperture of projected radius 0.2\arcsec, using a contiguous annulus 0.2\arcsec~in width for near-nucleus coma and background subtraction, finding $V$ = 20.09$\pm$0.01.  Given the observing geometry, and assuming a phase function correction of 0.04 magnitudes degree$^{-1}$, this corresponds to absolute magnitude $H$ = 14.42$\pm$0.01.    With assumed geometric albedo $p_V$ = 0.04, we compute a scattering cross-section $C = 6\times10^7$ m$^2$ within the 0.2\arcsec~aperture, corresponding to an  equal area circle of radius  4.4 km.  Since coma is present, this sets a strong upper limit to the radius of the nucleus.

 To determine a more stringent limit on the nucleus   we applied the convolutional surface brightness profile fitting method previously used on solar system comets (\cite{Lam98}, \cite{Hui18}).  This method represents the nucleus by a Dirac delta function, and uses a fit to the resolved portion of the coma to extrapolate inwards to the nucleus position, with convolution to account for the finite point spread function of the imaging system (HST has two pixel resolution $\sim$0.08\arcsec).  In the model, the ratio of the nucleus to coma signals is varied until the convolved model adequately matches the data.  Unfortunately, no convincing nucleus signal could be extracted from the HST data by this procedure, indicating the dominance of coma dust over nucleus scattering. \cite{Hui18} noted that the convolutional model works reliably only when the nucleus contributes $>$10\% of the total scattering cross-section because otherwise uncertainties in the spatial extrapolation of the coma profile dominate.  Using an aperture radius of 15 pixels (0.6\arcsec) for the coma fitting region we find a limit to the nucleus cross-section $C_n <  2.4\times10^7$  m$^2$, again assuming $p_V$ = 0.04.  This corresponds to a nucleus radius $r_n = (C_n/\pi)^{1/2} < 2.8$ km (corresponding to absolute magnitude $H >$15.4) and is our strongest constraint on the nucleus radius.  For comparison, the nuclei of 1I and 2I were approximately 0.1 km and 0.4 km in radius, respectively \citep{Jew23}.

\subsection{Morphology}
Visually, 3I is extended with respect to field stars (Figure \ref{image}).  At 3.8 au, gas is neither spectroscopically expected nor detected \citep{Opi25}, and  we can safely interpret the visible coma of 3I as due to scattering from ejected dust.  The dust is preferentially extended in a broad plume or fan along position angle 280$\pm$10\degr, about 180\degr~from the projected negative heliocentric velocity and anti-solar directions (Table \ref{geometry}).  Although described elsewhere as a ``sunward tail'' \citep{Cha25}, the emission to the north west is inconsistent with a tail produced by radiation pressure but  instead indicates anisotropic ejection towards the Sun.  Very faint emission is evident in the anti-solar direction, for which radiation pressure acceleration offers a natural explanation. (This radiation pressure tail is better seen in  strip-averaged perpendicular surface brightness profiles and is discussed later). Anisotropic mass loss is common in comets (e.g., \cite{Dor13}), where it is due to the preferential sublimation of ice on the hot day side of the nucleus and the near absence of sublimation on the night side.

\begin{figure}
\epsscale{0.9}

\plotone{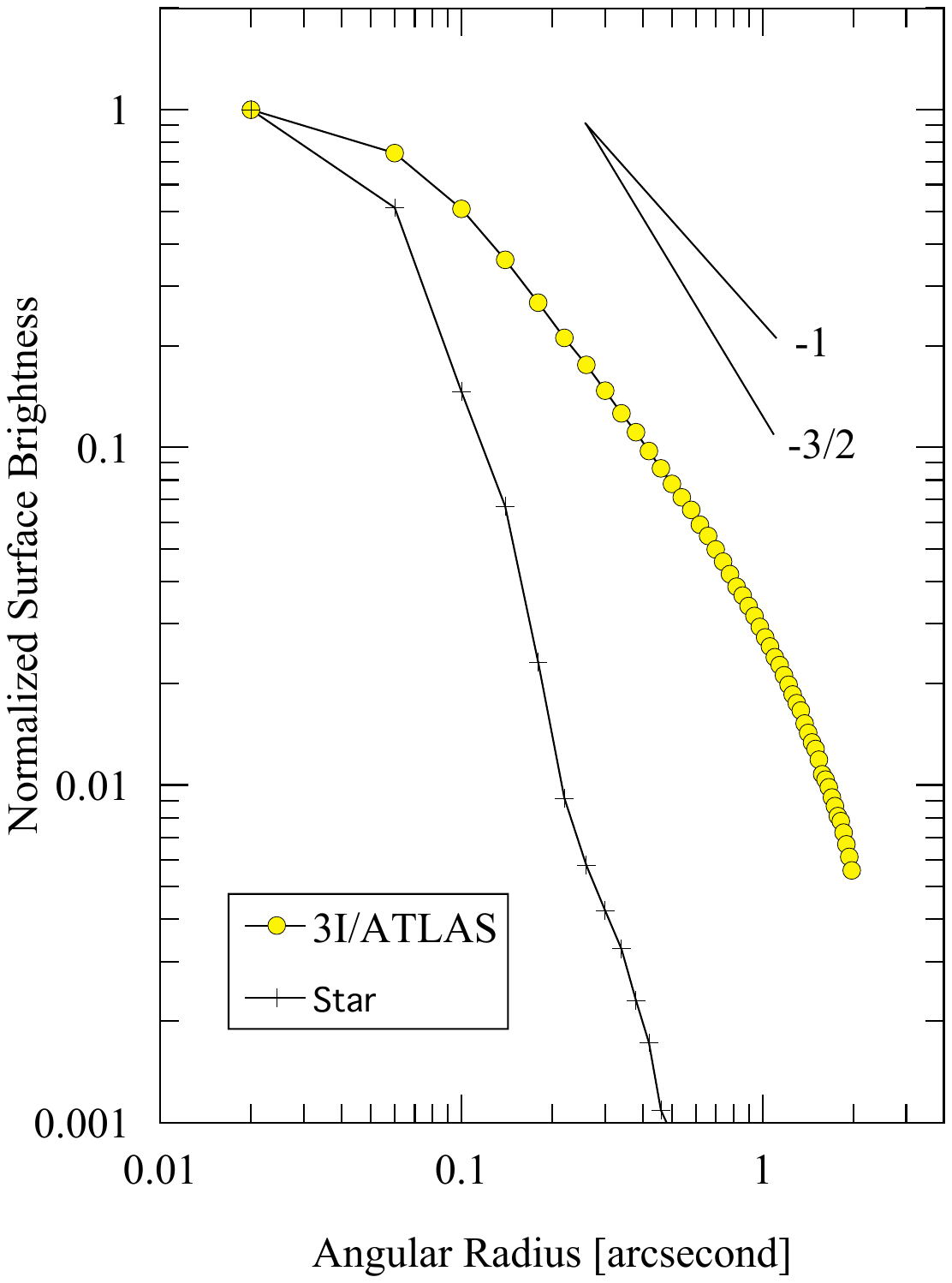}

\caption{Surface brightness plot computed using concentric circular annuli 1 pixel (0.04\arcsec) in width and centered on the photocenter.  The point spread function of HST is shown for comparison.  \label{SBplot}}
\end{figure}

The surface brightness, $\Sigma(p)$,  where $p$ is the angle from the center, was measured in a set of concentric annuli centered on the photocenter (Figure \ref{SBplot}).  The surface brightness follows $\Sigma(p) \propto p^{-m}$, with $m \sim$ 1 for angles $p <$0.4\arcsec, rising to $m \sim$1.5 at $p \sim$1\arcsec~and steepening to larger values as $p$ grows.  Gradient index $m$ = 1 is indicative of a coma expanding in steady state while the limiting case for dust accelerated by radiation pressure is $m$ = 1.5.  Steeper gradients suggest either that the  dust production rate had been quickly rising prior to the observation or that the grains are progressively destroyed as they flow away from the nucleus, causing a decrease in the surface brightness.  In this regard, a very weak band at 2 $\mu$m wavelength due to water ice has been reported in the coma of 3I \citep{Yan25}. It is possible that the ever steepening surface brightness profile in Figure \ref{SBplot} is a result of the progressive destruction of ice grains due to sublimation.  If so, we expect that the surface brightness profile should become steeper as the comet warms on approach to the Sun, at least until ice in the coma is completely exhausted.  

Coma morphology is influenced by many unmeasured parameters and is difficult to model even when in possession of much richer datasets than currently exist for 3I.  Nevertheless, several key parameters of the coma can be inferred from order-of-magnitude considerations, as follows. 
Specifically, we estimate dust parameters using first the  extent of the dust in the sunward direction and, second, the width of the radiation pressure swept tail.

We first consider a point source nucleus releasing dust in a broad distribution towards the Sun at speed $V_{\parallel}$.  The dust will experience a constant anti-sunward acceleration, causing it to slow down and reach a turning point at distance $X_R$ towards the Sun from the nucleus source.  The  magnitude of the acceleration is $\beta g_{\odot}$, where $g_{\odot}$ is the solar gravity and  $\beta$ is the radiation pressure efficiency factor.  For dielectric spheres, $\beta$ is related to the particle radius by $\beta \sim 1/a_{\mu}$, where $a_{\mu}$ is the radius expressed in microns \citep{Boh83}.  The dust speed and the turning distance are related by $V_{\parallel}^2 = 2 \beta g_{\odot} X_R$ \citep{Jew91}.  

In CCD data we can only  measure $x_R$, the projection of $X_R$ into the plane of the sky.  We assume $X_R = x_R/\sin(\alpha)$ where  $\alpha$ is the phase angle (c.f., Table \ref{geometry}).   Further writing $g_{\odot} = g_1/r_H^2$, where $g_1$ = 0.006 m s$^{-2}$ is the solar gravitational acceleration at 1 au and $r_H$ is the heliocentric distance in au, we obtain

\begin{equation}
V_{\parallel} = \left(\frac{2 \beta g_1 X_R}{ r_H^2 }\right)^{1/2}.
\label{nose}
\end{equation}

\begin{figure}
\epsscale{0.8}

\plotone{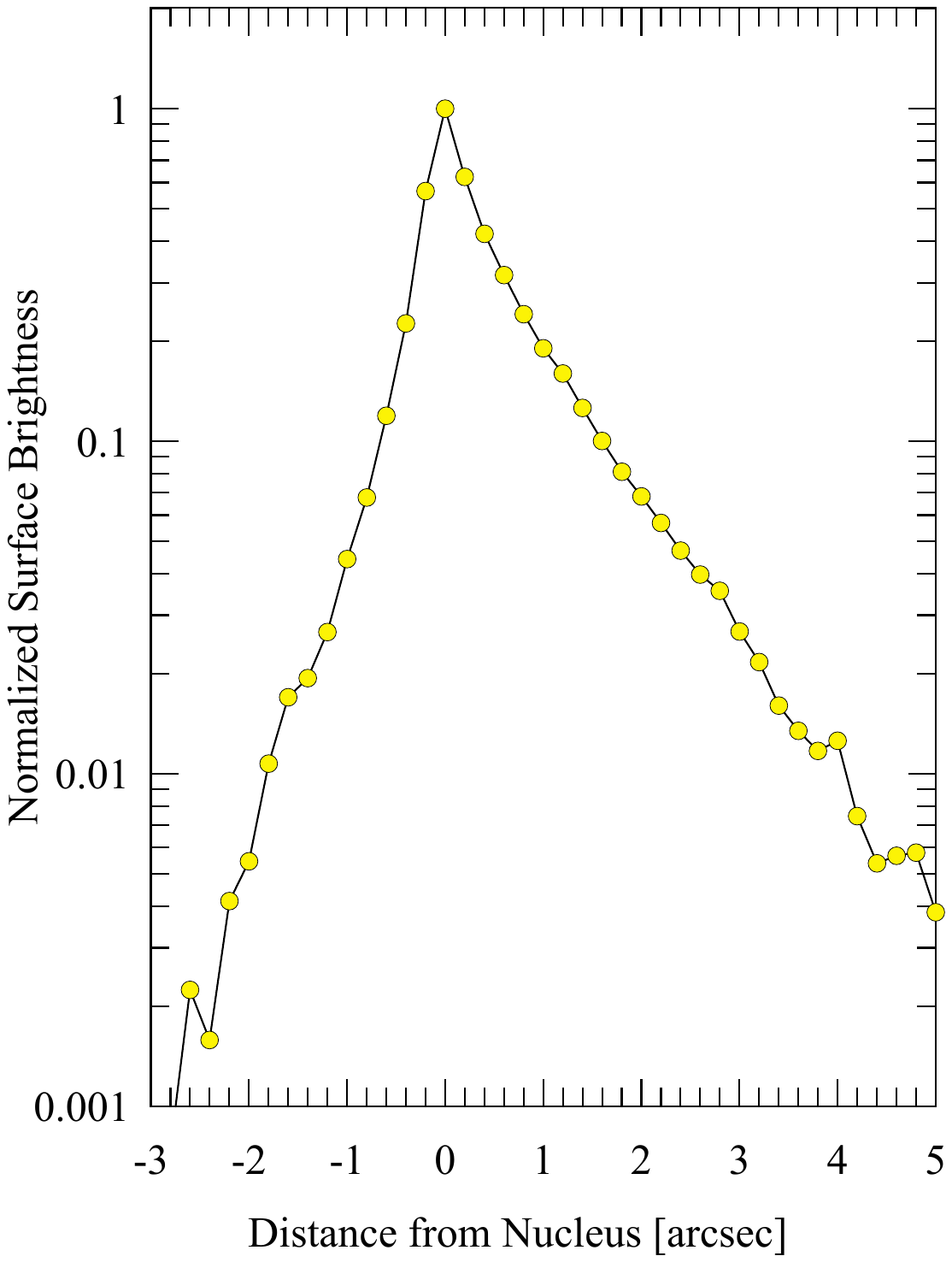}

\caption{Normalized surface brightness plotted as a function of $\ell$, the angular distance from the nucleus, with positive values being sunward.   The brightness was averaged over a 0.8\arcsec~wide strip along position angle 100\degr. \label{xplot}}
\end{figure}

\noindent Examination of the profile in Figure \ref{xplot}, measured in a 0.8\arcsec~wide strip along position angle 100\degr, shows that the surface brightness drops by an order of magnitude over a sky-plane distance $x_R$ = 1.5\arcsec, corresponding to $X_R = 3\times10^6$ m.  With $r_H$ = 3.83 au, Equation \ref{nose} gives 
$V_{\parallel}$ = 50 $\beta^{1/2}$ m s$^{-1}$ (equivalently, $V_{\parallel}$ = 50 $a_{\mu}^{-1/2}$).  A 1 $\mu$m particle ($\beta$ = 1) must be expelled at $V_{\parallel}$ = 50 m s$^{-1}$ while a 100 $\mu$m ($\beta \sim$0.01) particle would have $V_{\parallel} \sim$ 5 m s$^{-1}$ in order to reach  the sunward apex of the coma.  These speed estimates are predicated on the assumption that dust is primarily emitted from the nucleus of 3I in the sunward direction.

\begin{figure}
\epsscale{0.8}

\plotone{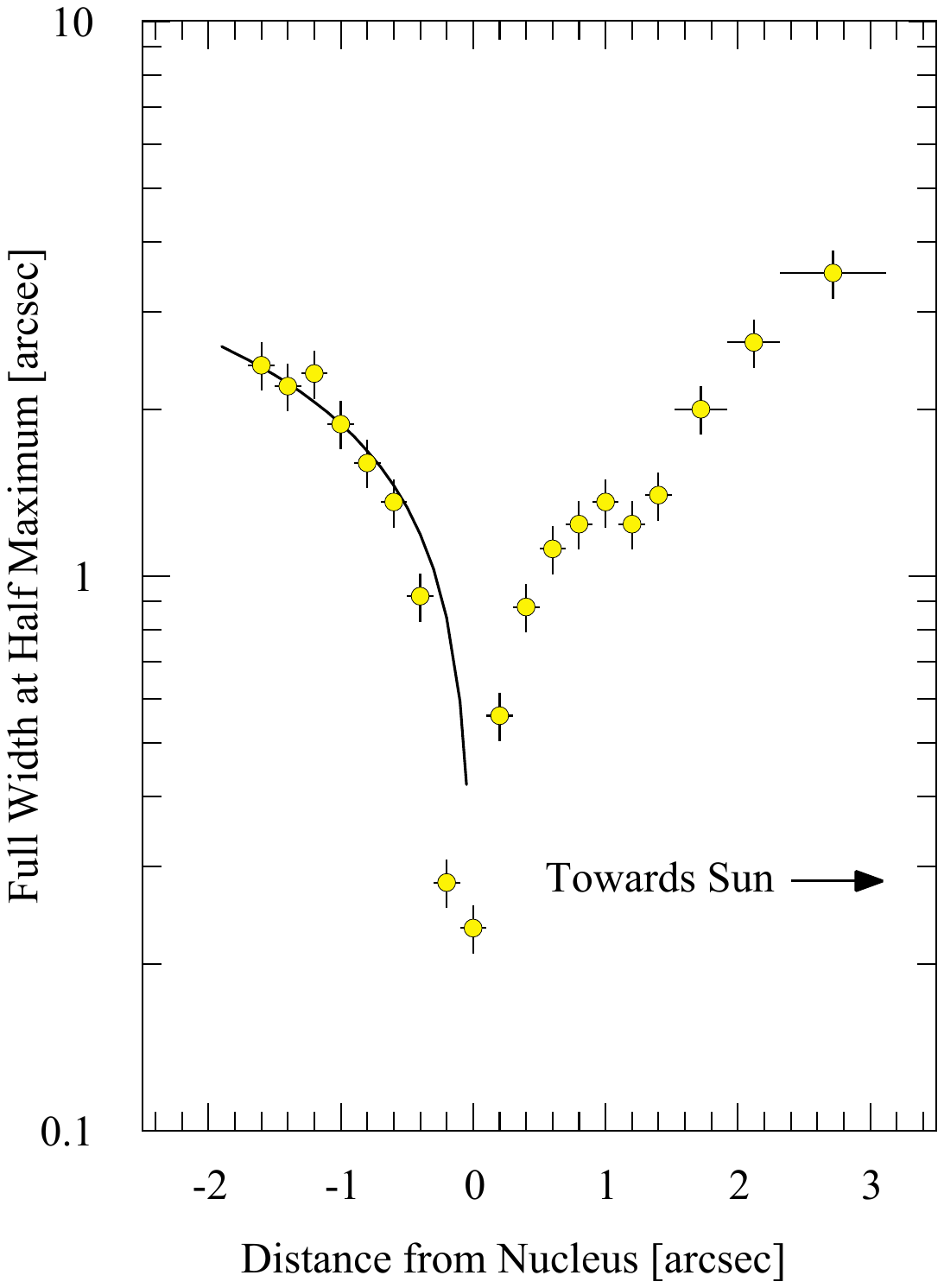}

\caption{Full width at half maximum measured perpendicular to the tail axis, as a function of $\ell$, the angular distance from the nucleus.  Horizontal bars indicate the widths over which the tail signal was binned.  The vertical bars indicate uncertainties, crudely estimated at $\pm$10\% of the FWHM.  The solid black line shows  FWHM $\propto |\ell|^{1/2}$, expected from the effects of radiation pressure acceleration on the tail particles. \label{yplot}}
\end{figure}

The width of the dust tail (swept to the East by radiation pressure) gives another particle constraint.  The FWHM of the tail measured perpendicular to its axis is plotted in Figure \ref{yplot}, where we have averaged over segments 0.2\arcsec~wide or more in order to improve the signal-to-noise ratio.  Clearly, the tail is broad, unlike the narrow trails observed in some comets \citep{Ish09} and in active asteroids (e.g., \cite{Jew14}). Narrow trails are produced by the release of large particles (often with $\beta$ = 10$^{-3}$ to 10$^{-2}$) at low speeds (typically $\lesssim$1 m s$^{-1}$).   

Given the edge-on observing geometry (c.f., Table \ref{geometry}) we can ignore the effects of projection on the vertical extent of the dust above the orbit plane. The distance traveled by a particle perpendicular to the orbit plane is simply $h = V_{\perp} t$, where $V_{\perp}$ is the perpendicular velocity and $t$ is the time elapsed since ejection.  The particle motion in the orbit plane, however, is accelerated by radiation pressure.  
As in \cite{Jew14}, we approximate the distance traveled in the plane in time $t$ by $L = \beta g_{\odot} t^2/2$ and eliminate $t$ between the equations for $h$ and $L$.  Again, to account for projection into the plane of the sky, we relate the distance traveled, $L$, to the sky-plane distance, $\ell$, using $L = \ell/\sin(\alpha)$.  Then

\begin{equation}
V_{\perp} = \left( \frac{\beta g_1 h^2 }{2 r_H^2 L}\right)^{1/2}.
\end{equation}

\noindent Substituting $\ell$ = 1\arcsec~($L= 1.3\times10^7$ m) and $h$ = 0.9\arcsec~(1.9$\times10^6$ m) we find $V_{\perp}$ = 7.5 $\beta^{1/2}$ m s$^{-1}$.  

Note that $V_{\parallel} > V_{\perp}$.  The sense of this inequality is readily understood as a consequence of preferential sublimation from the hot, Sun facing side of the nucleus.   Strong gas fluxes near and around the subsolar point accelerate dust to higher speeds than weaker, peripheral flow near the terminator or on the night side, producing the observed sunward fan.  The ratio  of the speeds gives an estimate of the sunward fan cone angle $\Theta \sim 2 \tan^{-1}(V_{\perp}/V_{\parallel}) \sim$ 17\degr. 

Small grains in near-Sun comets are easily accelerated by gas drag and commonly attain speeds comparable to $V_{th}$, the thermal speed of the sublimated gas molecules. We find that, in 3I, both $V_{\parallel}$ and $V_{\perp}$ are very small compared to $V_{th}$.  The thermal speed is

\begin{equation}
V_{th} = \left(\frac{8 k T}{\pi \mu m_H}\right)^{1/2}.
\label{temperature}
\end{equation}

\noindent where $k = 1.38\times10^{-23}$ J K$^{-1}$ is Boltzmann's constant, $T$ is the temperature of the sublimating ice, $\mu$ is the molecular weight and $m_H = 1.67\times10^{-27}$ kg is the mass of the hydrogen atom.  We solved the sublimation energy balance equation to find that, at 3.8 au, the temperature of a flat water ice surface ($\mu$ = 18) oriented normal to the Sun-comet line is $T$ = 183 K (i.e., slightly depressed relative to the 200 K temperature of an equivalent blackbody surface by the energy consumed in sublimation). Substitution in Equation \ref{temperature} gives  $V_{th} \sim$ 460 m s$^{-1}$.  The more volatile CO$_2$ and CO ices lead to larger temperature depression (as low as 27 K for CO) but still lead to thermal speeds $V_{th} > V_{\parallel}, V_{\perp}$.  Dust speeds much lower than $V_{th}$ indicate poor coupling, either because the particles are large, or because of weak gas flow (perhaps because sublimation occurs from beneath the physical surface of the nucleus through a porous mantle), or a source region of limited spatial extent \citep{Jew14} or some combination of these reasons.  While we do not currently possess sufficient data to decide between these possibilities, the faintness of the tail relative to the sunward ejected dust most simply suggests the dominance of large ($\beta \lesssim 10^{-2}$) particles.  We expect that a brighter tail should develop on approach to perihelion as large particles are progressively pushed back. As in 2I/Borisov and many solar system comets, small particles may be depleted in the coma as a result of inter-particle sticking forces.  Loss of small particles through sublimation might also play a role.


\subsection{Mass Loss Rate}
To estimate the mass loss rate in dust we use photometry within a 1\arcsec~radius circle, finding $V$ = 17.95, corresponding to $H$ = 12.47.  The scattering cross-section is related to $H$ by

\begin{equation}
C = \frac{2.25\times10^{22} \pi}{p_V}10^{-0.4(H - V_{\odot})}
\end{equation}

\noindent where $V_{\odot}$ = -26.74 is the V magnitude of the Sun at 1 au.  With $p_V$ = 0.04, we find the scattering cross-section, $C$, within 1\arcsec~as $C = 3.8\times10^8$ m$^2$.  For an optically thin assemblage of spherical particles of density $\rho$ and mean radius $a$, the total mass, $M$, and cross-section are related by $M = 4 \rho a C/3$.  We assume $\rho = 10^3$ kg m$^{-3}$ to find dust mass $M = 5.1\times10^5 a_{\mu}$ kg, where $a_{\mu}$ is the particle radius expressed in microns.  To maintain the dust population in steady state, this mass must be supplied on the aperture residence timescale, given by $t_r = d/V_{\parallel}$, where $d = 2.1\times10^6$ m is the radius of the 1\arcsec~photometry aperture projected to $\Delta$ = 2.98 au (Table \ref{geometry}).  We have used $V_{\parallel}$ rather than $V_{\perp}$ because the former better reflects the bulk of the mass loss from the hot day side of the nucleus (speeds and mass loss rates would be smaller by the ratio $V_{\perp}/V_{\parallel}$ if we instead assumed dust ejection at $V_{\perp}$).  Then, setting $dM/dt \sim M/t$ and using Equation \ref{nose}, we find

\begin{equation}
\frac{dM}{dt} = \frac{4 \rho \bar{a} C}{3d}\left(\frac{2 \beta g_1 X_R}{r_H^2}\right)^{1/2}
\end{equation}

\noindent Substituting values from above, and assuming the relation $\beta \sim 1/a_{\mu}$ for dielectric spheres, we find

\begin{equation}
\frac{dM}{dt} \sim 12~ a_{\mu}^{1/2}.
\label{dmbdt}
\end{equation}

\noindent With $a_{\mu}$ = 1, the implied steady state mass loss rate in dust is $dM/dt$ = 12 kg s$^{-1}$ and the aperture residence time is $t_r \sim$ 0.5 day.  Particles with $a_{\mu}$ = 100 would have $dM/dt$ = 120 kg s$^{-1}$ and $t_r \sim$ 5 day.  

A limit to the rate of production of the hydroxyl radical, OH, was spectroscopically set at $Q_{OH} < 8.2\times10^{26}$ s$^{-1}$ by \cite{Alv25} using data taken UT 2025 July 4 and 5 ($r_H \sim$ 4.4 au).  Ignoring the small fraction ($\sim$0.16) of water molecule dissociations that produce H$_2$ + O instead of H + OH, the corresponding mass loss rate in water molecules is $\mu m_H Q_{OH} <$ 25 kg s$^{-1}$.  We take the ratio of dust to gas production rates, $\psi$, as 0.5 $< \psi <$ 2 \citep{Mar25}.  With this range,  the OH data and Equation \ref{dmbdt},  taken together, indicate 1 $< a_{\mu} <$ 15, with a middle value ($\psi$ = 1) at $a_{\mu} \sim$ 4.  This estimate would be invalid if the driving volatile should turn out to be an ice other than water.

We solved the energy balance equation for the maximum sublimation rates at 3.8 au for water, carbon dioxide and carbon monoxide ices, finding $f_s \sim 1.2\times10^{-5}$ kg m$^{-2}$ s$^{-1}$, 1.4$\times10^{-4}$ kg m$^{-2}$ s$^{-1}$ and 3.1$\times10^{-4}$ kg m$^{-2}$ s$^{-1}$, respectively (see also Figure 1 of \cite{Jew25}). To supply $dM/dt$ = 12 $a_{\mu}^{1/2}$ kg s$^{-1}$ would require exposed areas of these ices $A_s = f_s^{-1} dM/dt$, corresponding to $A_s$ = 1.1 $a_{\mu}^{1/2}$ km$^{2}$, 0.09 $a_{\mu}^{1/2}$ km$^{2}$ and 0.04 $a_{\mu}^{1/2}$ km$^2$, respectively, setting lower limits to the diameter of a circular ice patch 2$(A_s/\pi)^{1/2} \sim$ 1.1 $a_{\mu}^{1/4}$ km, 0.34 $a_{\mu}^{1/4}$ km and 0.22 $a_{\mu}^{1/4}$ km, for these three ices.  For comparison, the nucleus of 2I/Borisov had effective circular radius in the range 0.2 km to 0.5 km, with a best estimate of $\sim$0.4 km. However, these are strong lower limits to the size of the 3I nucleus given that surface ice should be quickly depleted and that sublimation is likely to proceed from beneath the surface at lower specific rates over larger areas.  Moreover, in well-studied solar system comets, only a small fraction of the surface, typically 1\% to 10\%, sublimates \citep{Ahe95}, further emphasizing that considerations based on sublimation set a strong lower limit to the nucleus size.


\section{DISCUSSION}

Table \ref{comparison} offers a short comparison of the properties of the three known interstellar interlopers, with the parameters for 1I/'Oumuamua and 2I/Borisov being taken for convenience from the overview by \cite{Jew23}.  Mass loss rates in the Table are very approximate, are based on a variety of measurement techniques and models, and refer to data taken at different heliocentric distances.  To facilitate inter-comparison, we have scaled the rates to a common distance ($r_H$ = 1 au) assuming an $r_H^{-2}$ variation, which is the weakest likely heliocentric dependence.  The Table shows that the scaled mass loss from 3I is $>$10$^5$ to 10$^6$ times that from 1I/'Oumuamua and comparable to, or larger than, that from 2I/Borisov.  Judging by the mass loss rates, we suspect that the nucleus of 3I is a sub-kilometer body comparable in  radius to 2I/Borisov, and an order of magnitude smaller than early estimates.  Even so, the total mass loss from 3I at current rates is a negligible fraction of the total nucleus mass.

The high entry velocity of 3I suggests excitation by scattering around the galaxy for several to ten billion years \citep{Tay25}.   Irradiation by galactic cosmic rays over long periods in  interstellar space is expected to have damaged the molecular structure of the upper layers of the nucleus  \citep{Coo03}.  Cosmic ray damage  liberates hydrogen, which is mobile and escapes, leaving behind a high molecular weight, refractory ``irradiation mantle'' that would inhibit or prevent the sublimation of ices beneath.  While the molecular damage is strongest near the surface, high energy cosmic rays have stopping lengths of meters (density 10$^3$ kg m$^{-3}$ assumed) so that a considerable mass might be processed \citep{Joh91}. The observation that 3I is preferentially losing mass from its Sun-facing side suggests that such a protective irradiation mantle, if it exists, is thinner than the diurnal thermal skin depth.  For porous dielectric materials found on comets and other small solar system bodies (diffusivity $\kappa \sim$(1-10)$\times10^{-9}$  m$^2$ s$^{-1}$) and for comet-like rotation periods ($P \sim$ 10 to 15 hours), this skin depth is $(\kappa P)^{1/2} \sim$1 cm.  Alternatively, preferential day-side sublimation on 3I could reflect a  nucleus obliquity $\sim$90\degr~and a spin pole pointed approximately at the Sun.  In this case, continuous illumination of one hemisphere on the inbound leg would permit heat to penetrate a thicker irradiation mantle.   For example, in the two years since 3I crossed the 30 au orbit of Neptune (at which distance exposed CO ice is already volatile), heat could have penetrated a mantle $\sim$ 0.2 to 0.7 m thick, assuming the above diffusivity.  Future determination of the nucleus pole direction might be possible if rotationally modulated jets or other structures become apparent in the coma, allowing us to decide between these alternatives.

The absolute magnitude of the nucleus measured here ($H >$15.4) contrasts with ground-based measurements ($H$ = 12.3, \cite{Bol25}; $H \sim$ 12.4, \cite{Sel25} and \cite{Fei25}; $H$ = 13.7 ± 0.2, \cite{Cha25}) and proves the dominance of scattering from dust in 3I. Our finding that the optical cross-section is dominated by coma, not the nucleus, gives useful context for several observations  offered in recent preprints.  

First, the reddish color of 3I is consistent with the bulk colors of solar system comets (\cite{Alv25}, \cite{Bol25}, \cite{del25}, \cite{Opi25}, \cite{Sel25}, \cite{Yan25}), but this is a comparison only of dust colors in both 3I and the comets.  We have not measured the color of the nucleus, undercutting direct comparisons with Kuiper belt and other solar system bodies  and inhibiting attempts to assess the effects of prolonged interstellar exposure on its surface.  

Second, the dominance of coma should suppress any lightcurve due to the rotation of an underlying aspherical nucleus occurring on timescales less than the coma residence time \citep{Jew91}. As noted above,  the residence time is $t_r \sim 0.5 a_{\mu}^{1/2}$ day for 1\arcsec~apertures.   A modest photometric variation with a period 0.7 day was reported by \cite{del25}), but other photometry \citep{Sel25} shows instead a flat lightcurve.  For the moment, we must conclude that existing photometry of 3I  provides no sufficient basis on which to assess either the shape or the rotation of the nucleus of 3I.  Significantly, we cannot know from the existing data if its shape is highly aspherical, like that of 1I/'Oumuamua, or more nearly round, like typical small solar system bodies.

Third, and perhaps most significantly, the non-detection of the nucleus means that we cannot easily use the discovery of 3I to assess the number density of interstellar objects or their size distribution.  The HST upper limit to the nucleus radius does help alleviate (by a factor (2.8/10)$^3 \sim$ 0.02) the galactic mass budget problem caused by early over-estimates (10 km) of the radius (\cite{Sel25}, \cite{Loe25}).  \cite{Sel25} estimated the number density of objects with the scattering cross-section of 3I as $\sim3\times10^{-4}$ au$^{-3}$.  However,  coma around 3I complicates the interpretation of the ATLAS detection. 3I would have escaped detection by ATLAS if its cross-section and brightness had not been boosted by the dust coma. It is not obvious how to meaningfully scale the ATLAS detection (and the number density derived from it) to interloper nuclei of similar size but having less, or no, activity.  
For the time being, interloper population statistics must remain uncertain, tied as they are to 1I/'Oumuamua, the only inactive interloper detected to date.

\clearpage

\section{SUMMARY}

We present initial high resolution observations of interstellar interloper 3I/ATLAS from the Hubble Space Telescope.  The observations show that; 

\begin{itemize}

\item The nucleus radius determined from the surface brightness profile must be $r_n \le$ 2.8 km (absolute magnitude $H >$15.4, geometric albedo 0.04 assumed), and is likely much smaller.  

\item The optical scattering cross-section of 3I is dominated by dust ejected sunward, with a weaker anti-sun tail formed by radiation pressure (probably acting on large particles, radius $a \gg$ 1 $\mu$m).  There is no evidence for a narrow trail in the projected orbit, as would be formed by \textit{prolonged} release of large, slowly ejected particles.  

\item Mass loss rates in dust are $dM/dt = 12 a_{\mu}^{1/2}$ kg s$^{-1}$, where $a_{\mu}$ is the mean particle radius expressed in microns.  With $a_{\mu}$ in the range 1 to 100, dust mass loss rates are $dM/dt \sim$ 12 to 120 kg s$^{-1}$.
Dust ejection speeds are $V_{\parallel}$ = 50 $a_{\mu}^{-1/2}$ m s$^{-1}$ in the direction towards the Sun and $V_{\perp}$ = 7.5 $a_{\mu}^{-1/2}$ m s$^{-1}$  perpendicular to the orbit plane.  

\item Mass loss of the inferred order can be supplied by equilibrium sublimation from modest areas of exposed H$_2$O, CO$_2$ or CO ices. A strong minimum nucleus radius, $r_n \gtrsim$ 0.22$a_{\mu}^{1/4}$ km, is set by a model of the equilibrium production of coma by the sublimation of carbon monoxide, rising to 0.34$a_{\mu}^{1/4}$ km for the less volatile CO$_2$ ice or 1.1$a_{\mu}^{1/4}$ km for the least volatile H$_2$O.
\end{itemize}

\acknowledgments
We thank Darryl Seligman, Jane Luu and the anonymous referee for comments on the manuscript and Eric Keto for pointing out a binning error in the x-axis of Figure 3. Based on observations with the NASA/ESA Hubble Space Telescope
obtained [from the Data Archive] at the Space Telescope Science
Institute, which is operated by the Association of Universities for
Research in Astronomy, Incorporated, under NASA contract NAS5-
26555. Support for program number GO 17830 was
provided through a grant from the STScI under NASA contract NAS5-
26555.





\clearpage

\begin{deluxetable}{lcccrrrrrcrrrr}
\tabletypesize{\scriptsize}
\tablecaption{Observations 
\label{geometry}}
\tablewidth{0pt}
\tablehead{\colhead{UT Date\tablenotemark{a}} & \colhead{Filt,Exp\tablenotemark{b}} & \colhead{$r_H$\tablenotemark{c}}   & \colhead{$\Delta$\tablenotemark{d}} & \colhead{$\alpha$\tablenotemark{e}}  & \colhead{$\theta_{- \odot}$\tablenotemark{f}} & \colhead{$\theta_{-V}$\tablenotemark{g}} & \colhead{$\delta_{\oplus}$\tablenotemark{h}} & \colhead{$\nu$\tablenotemark{i}}     }

\startdata

July 21 16:29 - 16:32 & F350LP, 2$\times$40s & 3.830&  2.984&  9.6&  100.3&  97.4&  -0.67&  284.4\\
July 21 18:04 - 18:06 & F350LP, 2$\times$25s & 3.827&  2.983&  9.6&  100.2&  97.4&  -0.67&  284.4\\

\enddata


\tablenotetext{a}{UT Date in 2025 and start times of the first and last observation}
\tablenotetext{b}{Filter used and exposures obtained}
\tablenotetext{c}{Heliocentric distance, in au }
\tablenotetext{d}{Geocentric distance, in au }
\tablenotetext{e}{Phase angle, in degrees }
\tablenotetext{f}{Position angle of projected anti-solar direction, in degrees }
\tablenotetext{g}{Position angle of negative heliocentric velocity vector, in degrees}
\tablenotetext{h}{Angle of observatory from orbital plane, in degrees}
\tablenotetext{i}{True anomaly, in degrees}

\end{deluxetable}

\clearpage

\begin{deluxetable}{lcccccccrrrr}
\tablecaption{ Interlopers Compared\tablenotemark{a}
\label{comparison}}
\tablewidth{0pt}
\tablehead{\colhead{Object} & \colhead{$e$\tablenotemark{b}} & \colhead{$q$\tablenotemark{c}} & \colhead{$i$\tablenotemark{d}}   & \colhead{$r_n$\tablenotemark{e}} & \colhead{$a$\tablenotemark{f}} &\colhead{$dM/dt(r_H)$\tablenotemark{g}} & \colhead{$dM/dt(1)$\tablenotemark{h}}    }

\startdata
1I/'Oumuamua    & 1.201 & 0.256 & 122\degr.7 &  0.06-0.11 & N/A &$\lesssim10^{-3}$($\sim$1.4) & $\lesssim$2$\times10^{-3}$\\
2I/Borisov      & 3.356 & 2.006 & 44\degr.1  & 0.2-0.5    & $\sim$100 & 40(2.0-2.5) & 160-250\\
3I/ATLAS        & 6.145 & 1.357 & 175\degr.1 & $<$2.8     & $\sim$20? & 6-60(3.8) & 88-880\\

\enddata

\tablenotetext{a}{Data for 1I and 2I summarised from \cite{Jew23}}
\tablenotetext{b}{Orbital eccentricity}
\tablenotetext{c}{Perihelion distance, in au}
\tablenotetext{d}{Orbital inclination, in degrees }
\tablenotetext{e}{Radius of the nucleus represented as an equal area circle, in km}
\tablenotetext{f}{Effective particle radius, in $\mu$m}

\tablenotetext{g}{Measured mass loss rate(heliocentric distance of measurement), in kg s$^{-1}$ and au, respectively}
\tablenotetext{h}{Mass loss rate (kg s$^{-1}$) scaled to 1 au by $r_H^{-2}$}

\end{deluxetable}


\clearpage

\clearpage 

\clearpage 

\clearpage 

\end{document}